\documentclass[aps,prd,amsmath,amssymb,superscriptaddress,preprintnumbers,twocolumn,10pt,nofootinbib,showkeys]{revtex4-1}
\usepackage{graphicx}
\usepackage{dcolumn}
\usepackage{bm}
\usepackage{amssymb}
\usepackage{latexsym}
\usepackage{booktabs}
\usepackage{amsmath}
\usepackage{multirow}
\usepackage{url}
\usepackage{footnote}
\usepackage{float}
\usepackage{threeparttable}
\usepackage[colorlinks=true, linkcolor=blue, citecolor=blue]{hyperref}
\usepackage[bottom]{footmisc}

\usepackage[normalem]{ulem}
\usepackage{color}
\usepackage{array}
\usepackage{enumerate}
\usepackage{adjustbox}

\usepackage{makecell}
\usepackage{diagbox}
\usepackage{epstopdf}
\usepackage{epsfig}
\usepackage{longtable}
\usepackage{supertabular}
\usepackage{algorithm}
\usepackage{pifont}
\usepackage{algorithmic}
\usepackage{changepage}
\usepackage{setspace}
\newcommand{\diff}{\mathrm{d}}

\begin{document}

\title{Evidence for deviation in gravitational light deflection from general relativity at cosmological scales with KiDS-Legacy and CMB lensing}

\author{Guo-Hong Du}
\affiliation{Liaoning Key Laboratory of Cosmology and Astrophysics, College of Sciences, Northeastern University, Shenyang 110819, China}

\author{Tian-Nuo Li}
\affiliation{Liaoning Key Laboratory of Cosmology and Astrophysics, College of Sciences, Northeastern University, Shenyang 110819, China}

\author{Tonghua Liu}
\affiliation{School of Physics and Optoelectronic Engineering, Yangtze University, Jingzhou 434023, China}

\author{Jing-Fei Zhang}
\affiliation{Liaoning Key Laboratory of Cosmology and Astrophysics, College of Sciences, Northeastern University, Shenyang 110819, China}

\author{Xin Zhang}\thanks{Corresponding author}\email{zhangxin@neu.edu.cn}
\affiliation{Liaoning Key Laboratory of Cosmology and Astrophysics, College of Sciences, Northeastern University, Shenyang 110819, China}
\affiliation{Key Laboratory of Data Analytics and Optimization for Smart Industry (Ministry of Education), Northeastern University, Shenyang 110819, China}
\affiliation{National Frontiers Science Center for Industrial Intelligence and Systems Optimization, Northeastern University, Shenyang 110819, China}
 
\begin{abstract}

General relativity (GR) faces challenges from cosmic acceleration and observational tensions, necessitating stringent tests at cosmological scales. In this work, we probe GR deviations via a $\mu$--$\Sigma$ modified gravity parameterization, integrating KiDS-Legacy weak lensing (WL) data (1347 deg$^2$, $z\leq 2.0$), joint cosmic microwave background (CMB) data from Planck, ACT, and SPT, DESI DR2 baryon acoustic oscillation, and DES-Dovekie supernova data. KiDS-Legacy significantly improves constraint precision: $\mu_0$ (matter clustering) by $\sim 60\%$ and $\Sigma_0$ (gravitational light deflection) by $\sim 43\%$ relative to CMB alone. In the $\Lambda$CDM background, $\mu_0 = 0.21\pm 0.21$ is consistent with GR, while $\Sigma_0 = 0.149\pm 0.051$ deviates from GR at the 3.0$\sigma$ level. Furthermore, within the observationally preferred $w_0w_a$CDM background, this deviation in gravitational light deflection persists at the 2.2$\sigma$ level. This deviation is likely driven by the higher amplitudes in the large-scale CMB lensing measurements. This precise separation of GR-consistent matter clustering and deviant light deflection provides key observational clues for new physics or data systematics. Our work underscores the critical role of synergizing high-precision CMB and WL data in advancing GR tests.

\end{abstract}
\maketitle

\section{Introduction}\label{sec1}

General relativity (GR) constitutes the theoretical cornerstone of the standard cosmological model~\cite{Einstein:1915ca}. Since its inception, this theory has withstood extensive and precise tests ranging from solar system to cosmological scales~\cite{Berti:2015itd,Will:2014kxa}. However, the discovery of cosmic acceleration poses a severe challenge to this theoretical framework~\cite{SupernovaSearchTeam:1998fmf,SupernovaCosmologyProject:1998vns}. Within the GR framework, explaining this phenomenon typically requires introducing an energy component with negative pressure, namely dark energy (DE). The simplest candidate for DE is the cosmological constant $\Lambda$, which, together with cold dark matter, forms the $\Lambda$CDM model that has become the standard paradigm of contemporary cosmology.

Nevertheless, the cosmological constant faces profound theoretical difficulties, such as the fine-tuning and cosmic coincidence problems~\cite{Weinberg:1988cp,Sahni:1999gb,Carroll:2000fy,Peebles:2002gy}. On the observational front, the inconsistencies between early- and late-time measurements of the Hubble constant ($H_0$) and the amplitude of matter clustering ($S_8$), known as observational tensions, also imply potential defects in the standard model~\cite{Planck:2018vyg,DiValentino:2020vvd,DiValentino:2020zio,DiValentino:2021izs,DiValentino:2022fjm,Verde:2019ivm,Riess:2019qba,Riess:2021jrx,Giare:2023xoc,Guo:2018ans,Vagnozzi:2019ezj,Song:2022siz,Vagnozzi:2023nrq,Zhang:2024rra,Vagnozzi:2021gjh,Gao:2021xnk,Jin:2023sfc}. Consequently, the theoretical and observational issues associated with cosmic acceleration and DE necessitate further testing of GR at cosmological scales~\cite{Clifton:2011jh,Koyama:2015vza,Ishak:2018his}. This allows us to investigate whether cosmic acceleration is a derivative prediction within the GR framework (e.g., an unknown scalar field) or indicates that gravitational theory itself undergoes modification at infrared scales.

Over the past two decades, three main approaches have emerged for testing GR at cosmological scales based on observational data. The first approach involves analyzing specific self-consistent modified gravity (MG) theories or low-energy approximations of more fundamental theories, such as $f(R)$ gravity or the Dvali-Gabadadze-Porrati braneworld model~\cite{Carroll:2003wy,Capozziello:2003tk,Hu:2007nk,Dvali:2000rv,Clifton:2011jh,Li:2015poa,Langlois:2018dxi}. These theories garner attention due to their ability to recover GR at specific scales (e.g., solar system scales) via screening mechanisms. Although this approach offers a clear physical picture and explores the nature of gravity deeply, it is often limited by the computational complexity of deriving non-linear observables for each candidate theory. This complexity causes the development of this method in cosmological simulations to lag significantly behind that of the $\Lambda$CDM model~\cite{Alam:2020jdv}. The second approach is more indirect, focusing on quantifying inconsistencies among cosmological parameters derived from different datasets, interpreting these anomalies as issues within the standard model and its underlying gravitational theory, GR~\cite{DES:2020hen,Linder:2005in,DES:2020iqt,Ishak:2005zs}.

The third approach, adopted in this paper, introduces physically motivated phenomenological parametric modifications at the perturbation level of the Einstein equations~\cite{Clifton:2011jh,Koyama:2015vza}. In GR, these phenomenological parameters take specific values. Testing GR involves treating them as free parameters constrained by observational data to determine if they deviate from GR predictions. The advantage of this method lies in its ability to capture generic deviations from GR without assuming a specific MG model, thereby offering greater generality. Based on these results, one can further explore which specific models could generate such modifications. Typically, two phenomenological functions are employed to describe deviations from GR: $\mu(a,k)$, which modifies the Poisson equation for the Newtonian potential $\Psi$ affecting matter clustering, and $\Sigma(a,k)$ (or the gravitational slip $\eta = \Phi/\Psi$), which modifies the Poisson equation for the Weyl potential $(\Phi+\Psi)$ affecting light propagation. Consequently, these phenomenological effects are observationally most sensitive to matter clustering and gravitational lensing, with parameters closely related to $S_8$. Naturally, the background expansion history also influences these effects through parameter degeneracy.

Testing GR at cosmological scales necessitates high-precision survey data. Recently, the Dark Energy Spectroscopic Instrument (DESI) second data release (DR2) has precisely measured the cosmic expansion history using high-precision baryon acoustic oscillation (BAO) data derived from its 3-year observations~\cite{DESI:2025zgx}. When combined with cosmic microwave background (CMB) and type Ia supernova (SN) data (including the recalibrated DES-Dovekie samples), the results suggest a preference for dynamical DE at the 2.8--3.8$\sigma$ level within the $w_0w_a\mathrm{CDM}$ framework~\cite{DESI:2025zgx,DES:2025sig}. This deviation from the standard $\Lambda$CDM expansion history has sparked extensive and intense discussions regarding new physics~\cite{CosmoVerseNetwork:2025alb,Li:2024qso,Du:2024pai,RoyChoudhury:2024wri,Pedrotti:2025ccw,Jiang:2024viw,Jiang:2024xnu,Pedrotti:2024kpn,Giare:2024smz,Giare:2024gpk,Giare:2024oil,Giare:2025pzu,Colgain:2024xqj,Colgain:2025nzf,Fazzari:2025lzd,Li:2024qus,Li:2025owk,Ladeira:2026jne,Cheng:2025yue,Yang:2025uyv,Pan:2025qwy,Du:2025iow,Li:2025eqh,Li:2025dwz,Wang:2025znm,Huang:2025som,Wang:2026kbg,Wang:2024dka,Li:2025ops,Pang:2025lvh,Ye:2024ywg,Yao:2025kuz,Wu:2025vrl,Yao:2025twv,Wolf:2025jed,Wolf:2025acj,Li:2025msm,Song:2025bio,Silva:2025bnn,Ginat:2026fpo,Yin:2026gss,Li:2025vuh,Li:2025muv,Wu:2025vfs,Liu:2025myr,Wu:2025wyk,Liu:2025mub,Zhang:2025lam,Paliathanasis:2026ymi,deCruzPerez:2025dni,Alam:2025epg,Sabogal:2025qhz,Luciano:2025dhb,Luciano:2025ykr,Paliathanasis:2025kmg,Scherer:2025esj,Zhang:2025bmk,Schiavone:2026agq,Figueruelo:2026eis,Yang:2025gaz,Jia:2025poj,Wu:2024faw,Adam:2025kve,Zhou:2025nkb,vanderWesthuizen:2025rip,Wang:2025vtw,Li:2025ula,Li:2026xaz,Zhang:2025dwu,Du:2025xes,Cai:2025mas,Zheng:2025cgq,Schiavone:2026agq,Chaudhary:2025vzy,Zhang:2026cux}. Meanwhile, the DESI collaboration has also examined potential deviations from GR at the perturbation level at cosmological scales, utilizing the full shape (FS) modeling of clustering measurements based on the 1-year observations~\cite{Ishak:2024jhs}. The results indicate that current observational data remain consistent with GR across various parametrized models when combined with 3-year weak lensing data from Dark Energy Survey (DESY3) and CMB data (excluding Planck PR3 data\footnote{Numerous previous studies have indicated that Planck PR3 data exhibit an unphysical lensing parameter $A_{\mathrm{lens}}$ anomaly, which leads to lensing effects deviating abnormally from GR predictions; see, e.g., Refs.~\cite{Planck:2018vyg,Planck:2015bue,Ishak:2024jhs,Renzi:2017cbg,Mokeddem:2022bxa}. We explore the impact of $A_{\rm lens}$ and the total neutrino mass ($\sum m_\nu$) on our constraints in Appendix~\ref{appendixA}.}).

The latest weak lensing and CMB data provide new opportunities for testing GR. The Kilo-Degree Survey (KiDS) collaboration has released its final complete weak gravitational lensing measurement, KiDS-Legacy~\cite{Wright:2025xka,Reischke:2025hrt}. Covering 1347 $\mathrm{deg}^2$ with nine-band optical/near-infrared deep field imaging, KiDS-Legacy provides robust cosmic shear tomography up to $z \le 2.0$. Notably, through improved redshift calibration and image processing techniques, KiDS-Legacy has improved its constraining power on the structure amplitude $S_8$ by approximately 32\% compared to previous analyses. The measured value $S_8=0.815^{+0.016}_{-0.021}$ is almost consistent with Planck CMB observations (only a $0.73\sigma$ difference)~\cite{Wright:2025xka}. This finding provides key observational evidence for resolving the long-standing $S_8$ tension, although its robustness requires further verification. Meanwhile, the Atacama Cosmology Telescope (ACT) and the South Pole Telescope (SPT) collaborations recently jointly released small-scale CMB anisotropy measurements~\cite{AtacamaCosmologyTelescope:2025blo,SPT-3G:2025bzu} and the most precise lensing reconstruction data derived from ACT, SPT, and Planck experiments~\cite{ACT:2025qjh}. CMB lensing probes the integrated Weyl potential to the last scattering surface, forming a complementarity with the low-redshift galaxy weak-lensing effects probed by KiDS. Therefore, integrating these latest high-precision cosmological datasets to effectively constrain phenomenological properties of gravity is both important and of significant research value.

In this work, we utilize the latest joint CMB data from Planck, ACT, and SPT, weak lensing data from KiDS-Legacy, as well as DESI DR2 BAO data and DES-Dovekie SN data to test GR at cosmological scales. At the background level, we consider both $\Lambda$CDM and $w_0w_a\mathrm{CDM}$ evolutions, while at the perturbation level, we consider a typical $\mu$-$\Sigma$ parametrized MG model. We report the parameter constraint results, perform a detailed analysis of the data fit, and emphasize the significant role of high-precision lensing data in testing GR.

This paper is organized as follows. In Sec.~\ref{sec2}, we briefly introduce the parametrized models considered and the cosmological data used in the analysis. In Sec.~\ref{sec3}, we report the constraint results and provide relevant discussions. The conclusion is given in Sec.~\ref{sec4}.

\section{Theoretical Framework} \label{sec2}

We examine any deviations from the predictions of GR by introducing phenomenological parameters into the perturbed Einstein field equations and constraining them using observational data~\cite{Clifton:2011jh,Koyama:2015vza}. We adopt the context of a spatially flat Friedmann-Lema\^{i}tre-Robertson-Walker metric. Considering scalar perturbations in the conformal Newtonian gauge, the line element is given by
\begin{equation}
\diff s^2 = a^2(\tau) \left[ -(1+2\Psi)\diff\tau^2 + (1-2\Phi)\delta_{ij}\diff x^i \diff x^j \right],
\label{eq:metric}
\end{equation}
where $\tau$ is the conformal time, $a(\tau)$ is the scale factor normalized to unity at the present day, and $\Psi$ and $\Phi$ are the gauge-invariant scalar potentials. Physically, $\Psi$ corresponds to the Newtonian potential governing the motion of non-relativistic particles (clustering), while $\Phi$ represents the perturbation to the spatial curvature.

In GR, the Einstein field equations relate the metric perturbations to the energy-momentum tensor. For a fluid with density perturbation $\delta$ and negligible anisotropic stress (a valid assumption for non-relativistic matter and DE in the standard model), the Einstein equations in Fourier space yield two key equations. First, the Poisson equation is given by
\begin{equation}
k^2 \Psi = -4\pi G a^2 \bar{\rho} \Delta,
\label{eq:poisson_gr}
\end{equation}
where $\bar{\rho}$ is the background matter density and $\Delta$ is the comoving density contrast. Second, the difference between the two potentials is sourced by the anisotropic stress $\sigma$ of the cosmic fluid, described by the relation,
\begin{equation}
k^2 (\Psi - \Phi) = -12\pi G a^2 (\bar{\rho} + \bar{P}) \sigma.
\end{equation}
Since the contribution from relativistic species (which possess non-zero shear stress) is negligible at late cosmological epochs, and the anisotropic stress vanishes for non-relativistic matter and DE, this equation implies the gravitational slip vanishes, yielding $\Psi = \Phi$.

To test for deviations from GR, we introduce two phenomenological functions, $\mu(a,k)$ and $\Sigma(a,k)$, which modify the relations between the metric potentials and the matter density field~\cite{Clifton:2011jh,Koyama:2015vza}. The modified Poisson equation is defined as
\begin{equation}
k^2 \Psi = -4\pi G a^2 \mu(a,k) \bar{\rho} \Delta.
\label{eq:mu_def}
\end{equation}
The function $\mu(a,k)$ effectively rescales the gravitational constant $G_{\rm eff} = \mu(a,k) G$ felt by massive particles. This parameter is directly probed by galaxy clustering and redshift-space distortion measurements.

Gravitational lensing and the Integrated Sachs-Wolfe effect are sensitive to the sum of the potentials, $(\Phi+\Psi)$, often referred to as the Weyl potential. We adopt parameterized deviations in this sector using $\Sigma(a,k)$,
\begin{equation}
k^2 (\Phi + \Psi) = -8\pi G a^2 \Sigma(a,k) \bar{\rho} \Delta.
\label{eq:sigma_def}
\end{equation}
The parameter $\Sigma(a,k)$ describes modifications to the propagation of light. An alternative parametrization uses the gravitational slip $\eta(a,k) = \Phi/\Psi$. The relationship between these parameters is given by $\Sigma = \frac{\mu}{2}(1+\eta)$. In GR, $\mu = \Sigma = \eta = 1$.

Since the functional dependencies of $\mu$ and $\Sigma$ on scale $k$ and time $a$ are unknown a priori, we must adopt specific forms to perform a likelihood analysis. We focus on the ``late-time'' modification scenario, motivated by the coincidence of cosmic acceleration and the onset of DE dominance. For modifications that are scale-independent on sub-horizon scales (valid for the quasi-static regime of many scalar-tensor theories), we adopt a parameterization proportional to the DE density $\Omega_{\rm DE}(a)$, given by~\cite{Planck:2015bue}
\begin{align}
\mu(a) &= 1 + \mu_0 \frac{\Omega_{\rm DE}(a)}{\Omega_\Lambda}, \label{eq:mu_time} \\
\Sigma(a) &= 1 + \Sigma_0 \frac{\Omega_{\rm DE}(a)}{\Omega_\Lambda}, \label{eq:sigma_time}
\end{align}
where $\Omega_\Lambda$ is the DE density parameter at $z=0$. The free parameters $\mu_0$ and $\Sigma_0$ quantify the amplitude of the deviation today. In GR, $\mu_0 = \Sigma_0 = 0$.

For the background evolution, we perform our analysis in two background expansion scenarios:
\begin{itemize}
\item $\boldsymbol{\Lambda \mathrm{CDM}}$: The equation of state (EoS) of DE is fixed to $w = -1$. Deviations from GR manifest purely in the perturbations. When introducing MG parameters, the model is labelled as $\boldsymbol{\mu_0\Sigma_0\Lambda\mathrm{CDM}}$.
\item $\boldsymbol{w_0w_a\mathrm{CDM}}$: We allow for a time-evolving DE EoS using the Chevallier-Polarski-Linder parametrization~\cite{Chevallier:2000qy, Linder:2002et}, given by
\begin{equation}
w(a) = w_0 + w_a (1-a).
\end{equation}
This is particularly relevant given recent results from DESI DR2 indicating a preference for dynamical DE ($w_0 > -1, w_a < 0$)~\cite{DESI:2025zgx}. When introducing MG parameters, the model is labelled as $\boldsymbol{\mu_0\Sigma_0w_0w_a}$.
\end{itemize}

\begin{table*}[!htb]
\renewcommand\arraystretch{1.5}
\centering
\caption{Cosmological parameter constraints for the $\Lambda$CDM, $w_0 w_a$CDM, $\mu_0 \Sigma_0 \Lambda$CDM, and $\mu_0 \Sigma_0 w_0 w_a$ models, utilizing CMB, DESI, DES-Dovekie, and KiDS-Legacy data. Note that ``CMB-nl'' denotes the exclusion of CMB lensing data. Here, $H_0$ is in units of $\rm km\,s^{-1}\,Mpc^{-1}$.}
\label{table1}
\resizebox{\textwidth}{!}{
\setlength{\tabcolsep}{5pt}
\begin{tabular}{lcccccc}
\hline\hline
Model / Data & $H_0$ & $S_8$ & $w_0$ & $w_a$ & $\mu_0$ & $\Sigma_0$ \\
\hline

\multicolumn{7}{l}{$\boldsymbol{\Lambda\mathrm{CDM}}$} \\
CMB+DESI+DES-Dovekie             & $68.03\pm 0.24$ & $0.820\pm 0.006$ & $\text{---}$ & $\text{---}$ & $\text{---}$ & $\text{---}$ \\
CMB+DESI+DES-Dovekie+KiDS-Legacy & $68.02\pm 0.22$ & $0.819\pm 0.005$ & $\text{---}$ & $\text{---}$ & $\text{---}$ & $\text{---}$ \\
\hline

\multicolumn{7}{l}{$\boldsymbol{w_0w_a\mathrm{CDM}}$} \\
CMB+DESI+DES-Dovekie             & $67.45\pm 0.53$ & $0.833\pm 0.007$ & $-0.802\pm 0.055$ & $-0.75\pm 0.21$ & $\text{---}$ & $\text{---}$ \\
CMB+DESI+DES-Dovekie+KiDS-Legacy & $67.43\pm 0.53$ & $0.830\pm 0.006$ & $-0.801\pm 0.054$ & $-0.74\pm 0.20$ & $\text{---}$ & $\text{---}$ \\
\hline

\multicolumn{7}{l}{$\boldsymbol{\mu_0\Sigma_0\Lambda\mathrm{CDM}}$} \\
CMB-nl                           & $67.83\pm 0.52$ & $0.831^{+0.061}_{-0.085}$ & $\text{---}$ & $\text{---}$ & $0.11^{+0.57}_{-0.80}$ & $0.154^{+0.099}_{-0.110}$ \\
CMB                              & $67.75\pm 0.48$ & $0.837^{+0.064}_{-0.082}$ & $\text{---}$ & $\text{---}$ & $0.16^{+0.61}_{-0.75}$ & $0.112^{+0.086}_{-0.100}$ \\
CMB+DESI+DES-Dovekie             & $68.31\pm 0.26$ & $0.830^{+0.065}_{-0.092}$ & $\text{---}$ & $\text{---}$ & $0.23\pm 0.67$ & $0.146\pm 0.094$ \\
CMB+KiDS-Legacy                  & $67.81\pm 0.46$ & $0.818\pm 0.021$ & $\text{---}$ & $\text{---}$ & $0.02\pm 0.27$ & $0.130\pm 0.053$ \\
CMB+DESI+DES-Dovekie+KiDS-Legacy & $68.32\pm 0.27$ & $0.826\pm 0.019$ & $\text{---}$ & $\text{---}$ & $0.21\pm 0.21$ & $0.149\pm 0.051$ \\
\hline

\multicolumn{7}{l}{$\boldsymbol{\mu_0\Sigma_0w_0w_a}$} \\
CMB+DESI+DES-Dovekie             & $67.41\pm 0.57$ & $0.833^{+0.056}_{-0.081}$ & $-0.824\pm 0.056$ & $-0.60^{+0.23}_{-0.20}$ & $0.13^{+0.52}_{-0.93}$ & $0.107^{+0.076}_{-0.087}$ \\
CMB+KiDS-Legacy                  & $70.9^{+5.40}_{-6.50}$ & $< 0.870\,(2\sigma)$ & $-0.800^{+0.290}_{-0.400}$ & $-1.23^{+0.51}_{-1.80}$ & $0.18\pm 0.44$ & $0.095^{+0.061}_{-0.069}$ \\
CMB+DESI+DES-Dovekie+KiDS-Legacy & $67.44\pm 0.54$ & $0.826\pm 0.021$ & $-0.827\pm 0.055$ & $-0.59\pm 0.21$ & $0.09^{+0.28}_{-0.24}$ & $0.115\pm 0.053$ \\
\hline
\end{tabular}
}
\end{table*}

We employ the publicly available sampler \texttt{Cobaya}~\cite{Torrado:2020dgo} to perform Markov Chain Monte Carlo (MCMC) analyses for constraining cosmological parameters. Theoretical predictions are computed using an extended nonlinear version of \texttt{MGCAMB}~\cite{Zucca:2019xhg,Wang:2023tjj}, which incorporates various MG parameterization schemes. Within this extended framework, the \texttt{ReACT} code~\cite{Bose:2020wch,Bose:2022vwi} is utilized to extend the calculated linear theoretical power spectra into the nonlinear regime to consistently model the nonlinear matter power spectrum under the modifications of $\mu$ and $\Sigma$. For the CMB, BAO, and SN likelihoods, we utilize both the internal implementations within \texttt{Cobaya} and the available external likelihoods. Regarding the KiDS-Legacy likelihood, for which the standard library originates from \texttt{CosmoSIS}~\cite{Zuntz:2014csq}, we employ the \texttt{cosmosis2cobaya} interface for integration~\cite{Ye:2024rzp}. We assess the convergence of the chains using the Gelman-Rubin statistic $R-1 < 0.02$~\cite{Gelman:1992zz} and analyze the resulting MCMC chains using the \texttt{GetDist} package~\cite{Lewis:2019xzd}. The datasets employed in this work are summarized below:
\begin{itemize}

\item \textbf{\texttt{CMB}:} We utilize Planck 2018 CMB temperature (TT), E-mode polarization (EE), and cross-correlation (TE) spectra data modeled via the \texttt{Commander}, \texttt{SimAll}, and \texttt{CamSpec} likelihoods~\cite{Efstathiou:2019mdh,Rosenberg:2022sdy}. In addition to the Planck data, we utilize the high-multipole CMB measurements from ACT DR6~\cite{AtacamaCosmologyTelescope:2025blo} and SPT-3G~\cite{SPT-3G:2025bzu}. To strictly avoid mode double-counting, the Planck CMB spectra are restricted to $\ell < 1000$ for TT and $\ell < 600$ for TE and EE, respectively. Finally, we incorporate the joint CMB lensing power spectrum reconstructed from Planck PR4, ACT DR6, and SPT-3G maps~\cite{ACT:2025qjh,Carron:2022eyg,ACT:2023dou,ACT:2023kun}.

\item \textbf{\texttt{KiDS-Legacy}:} We incorporate the weak lensing measurements from the final KiDS release, KiDS-Legacy~\cite{Wright:2025xka}. This dataset represents the most precise and robust analysis from KiDS to date, covering 1347 deg$^2$ with nine-band optical/near-infrared imaging. A primary advantage of KiDS-Legacy is its significantly extended redshift baseline ($z \le 2.0$), enabled by enhanced spectroscopic calibration and deep near-infrared photometry, which effectively breaks parameter degeneracies~\cite{Wright:2025xka}. Furthermore, rigorous improvements in redshift distribution estimation and shape calibration have substantially suppressed systematic uncertainties, resulting in an approximate 32\% increase in constraining power on the structure growth amplitude compared to previous KiDS analyses~\cite{Wright:2025xka}.

\item \textbf{\texttt{DESI}:} The BAO constraints utilized in this analysis are derived from the DESI DR2 observations of galaxies, quasars, and the Lyman-$\alpha$ forest, as tabulated in Table IV of Ref.~\cite{DESI:2025zgx}. These data provide measurements of the transverse comoving distance $D_{\mathrm{M}}/r_{\mathrm{d}}$, the angle-averaged distance $D_{\mathrm{V}}/r_{\mathrm{d}}$, and the Hubble horizon $D_{\mathrm{H}}/r_{\mathrm{d}}$, with all distances normalized by the comoving sound horizon at the drag epoch, $r_{\mathrm{d}}$.

\item \textbf{\texttt{DES-Dovekie}:} This analysis incorporates the DES-Dovekie dataset, a reconstructed compilation resulting from a comprehensive re-calibration and re-analysis of the DESY5 SN sample~\cite{DES:2025sig}. The catalog comprises a total of 1820 supernovae, consisting of 1623 probable candidates (defined by $P_{\mathrm{SNe}} > 0.5$) combined with 197 low-redshift supernovae.

\end{itemize}

\section{Results and Discussion}\label{sec3}

\begin{figure*}[htbp]
\includegraphics[scale=0.45]{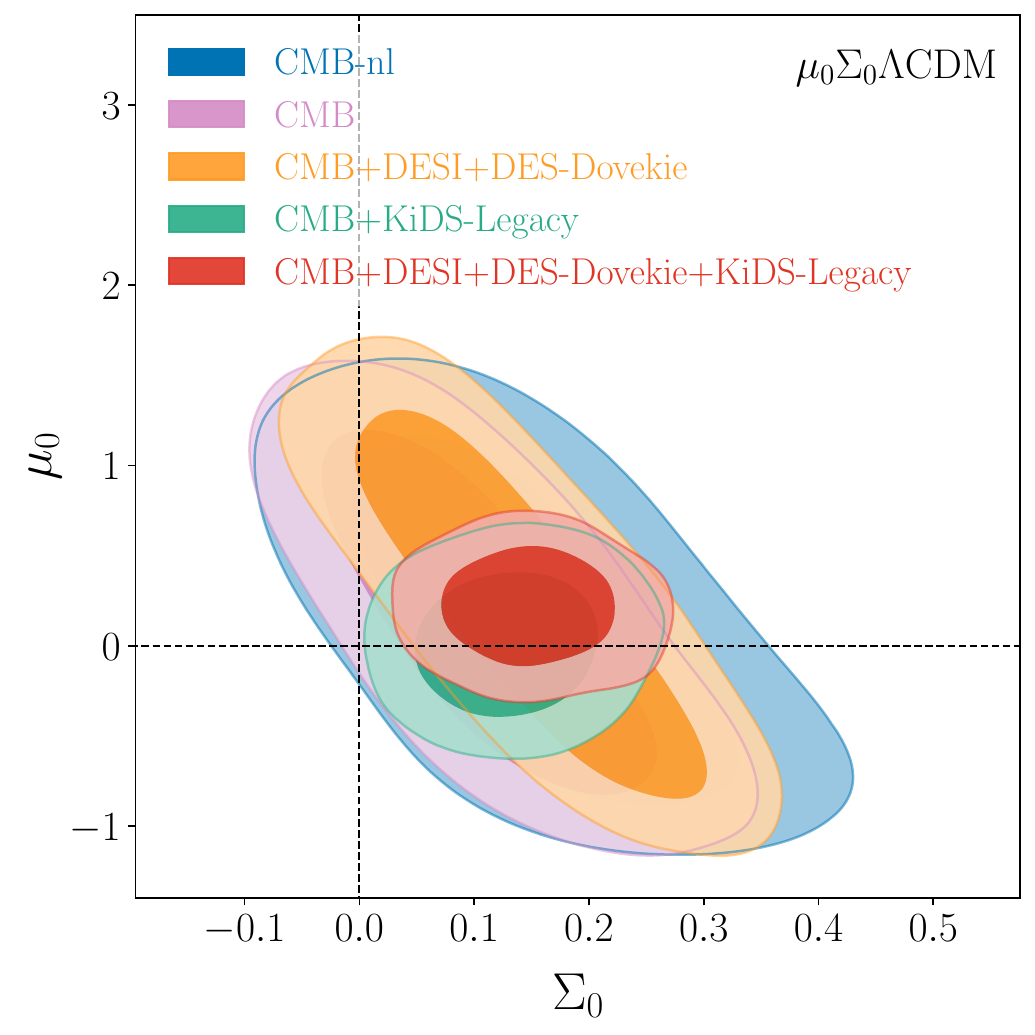}
\includegraphics[scale=0.45]{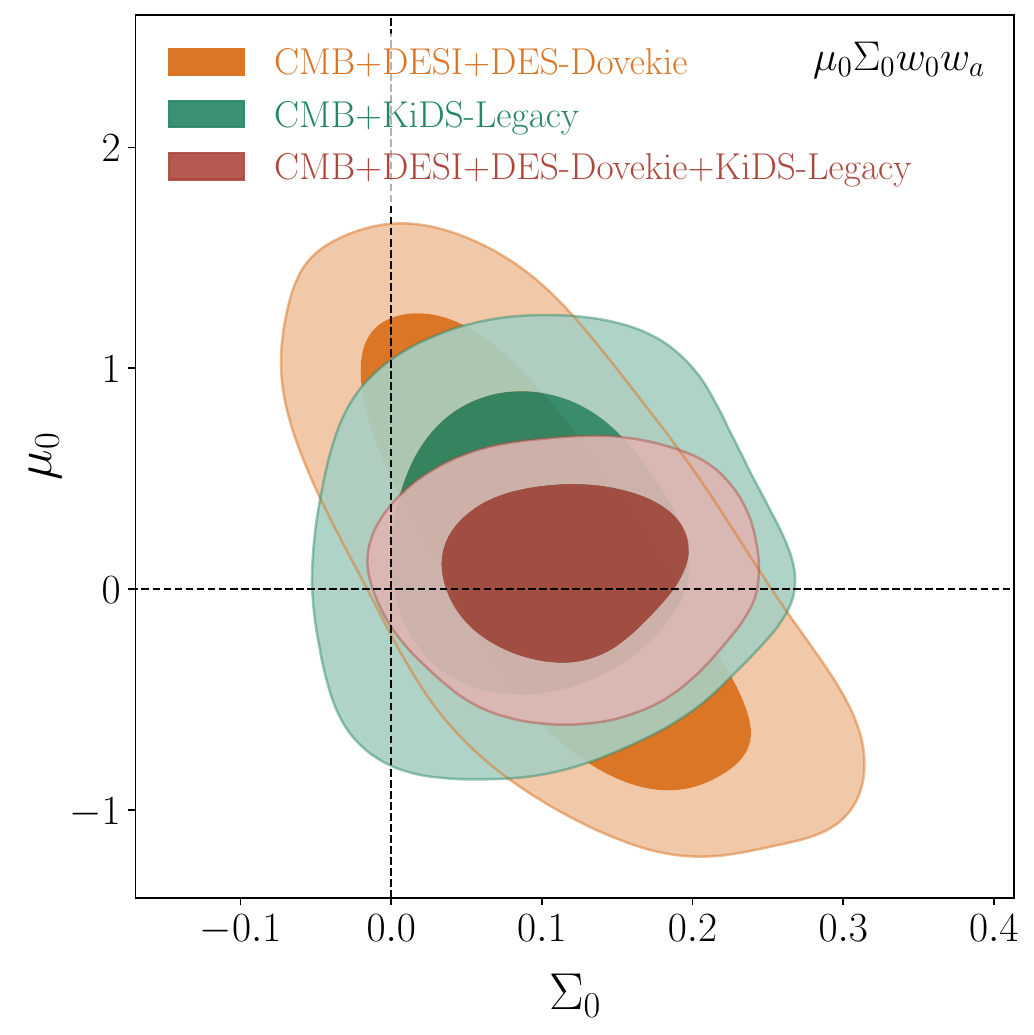}
\centering
\caption{\label{fig1} The $1\sigma$ and $2\sigma$ credible-interval contours for the $\mu_0$ and $\Sigma_0$ parameters, utilizing the CMB, DESI, DES-Dovekie, and KiDS-Legacy data. The results for the $\mu_0 \Sigma_0 \Lambda$CDM and $\mu_0 \Sigma_0 w_0 w_a$ models are displayed in the left and right panels, respectively.}
\end{figure*}

\begin{figure*}[htbp]
\includegraphics[scale=0.48]{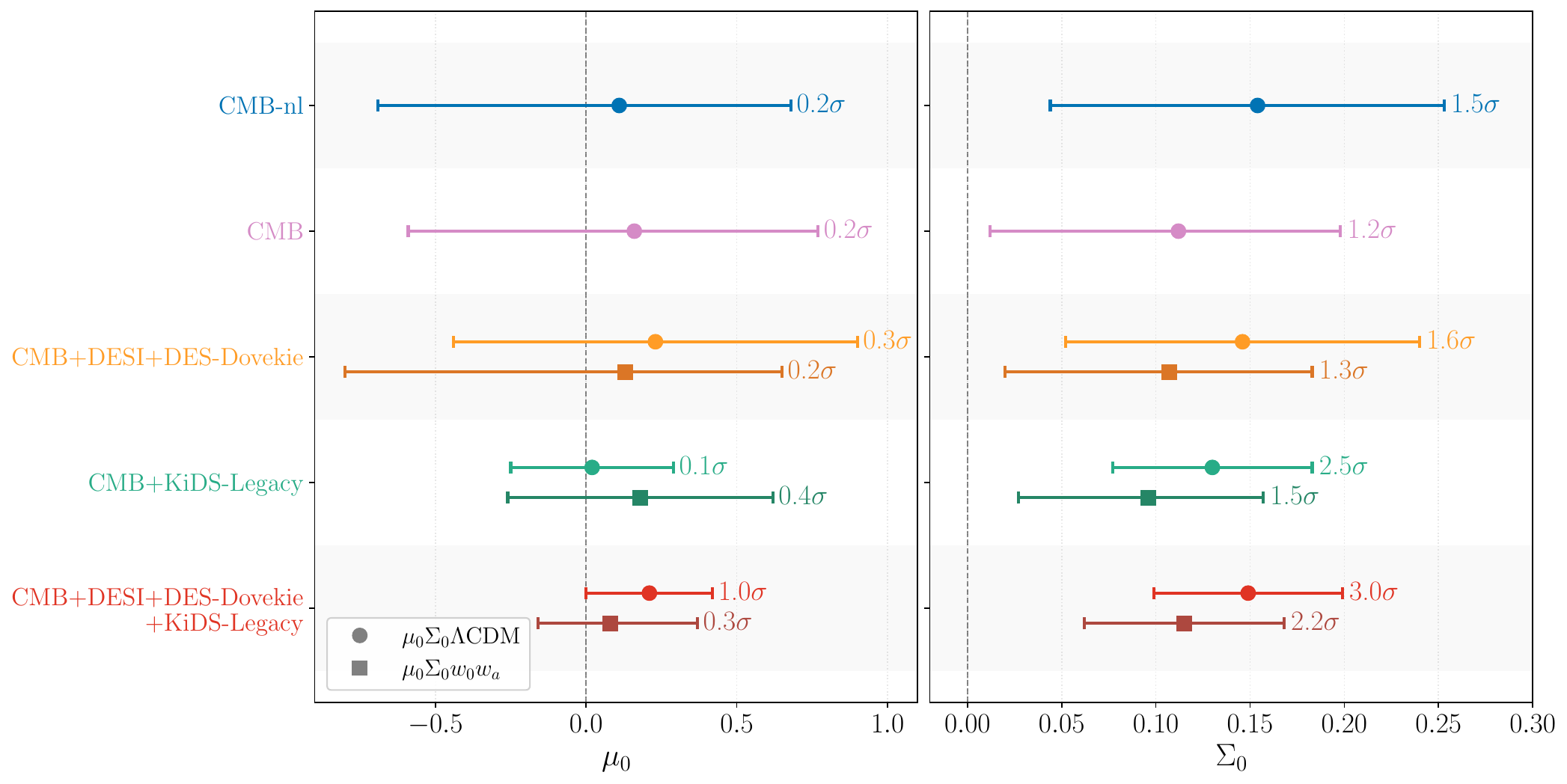}
\centering
\caption{\label{fig2} Whisker plots for the $\mu_0$ and $\Sigma_0$ parameters using the CMB, DESI, DES-Dovekie, and KiDS-Legacy data. The results for the $\mu_0 \Sigma_0 \Lambda$CDM and $\mu_0 \Sigma_0 w_0 w_a$ models are represented by circles and squares, respectively.}
\end{figure*}

\begin{figure*}[htbp]
\includegraphics[scale=0.75]{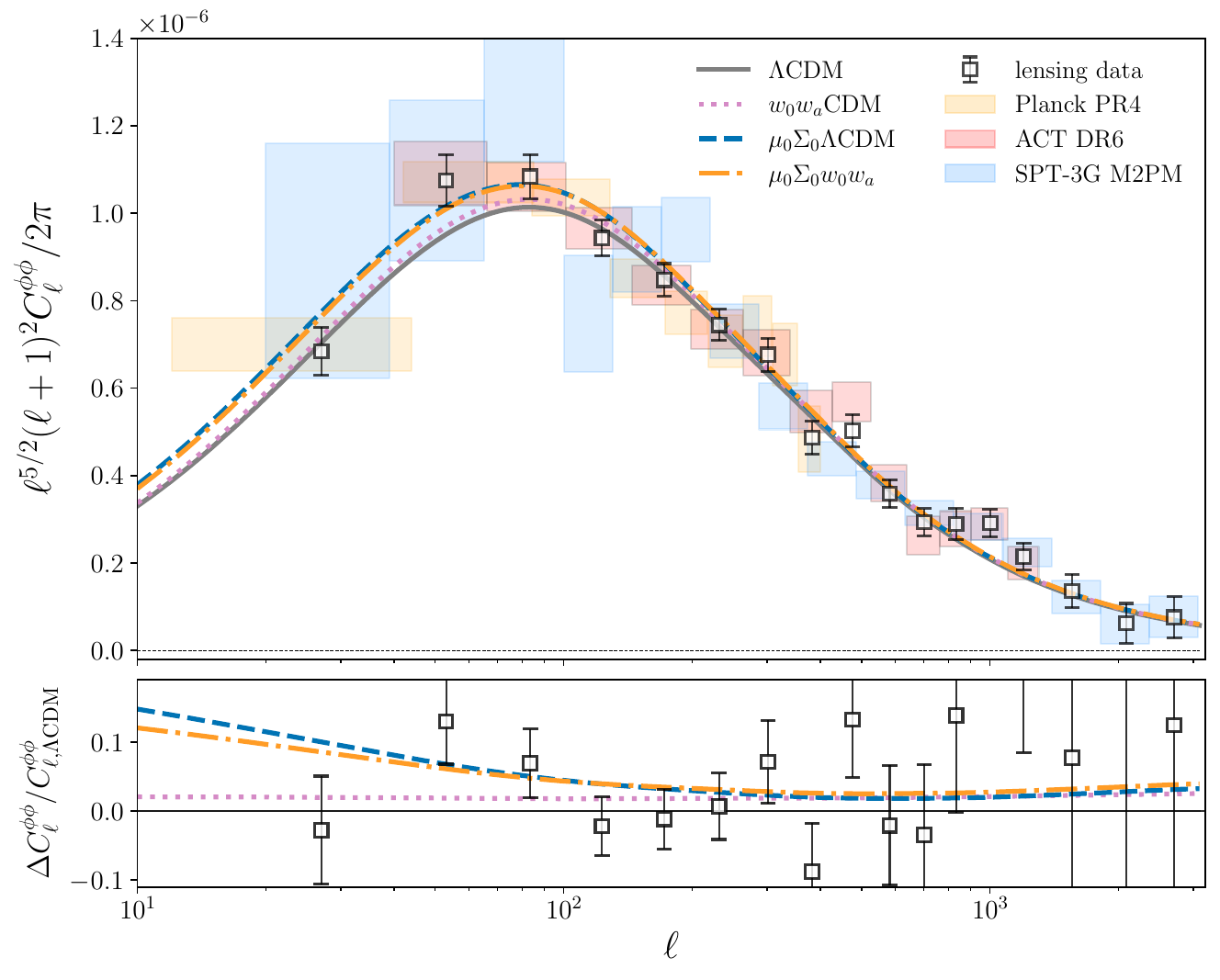}
\centering
\caption{\label{fig3} The CMB lensing potential power spectrum (upper panel) and fractional residuals relative to the best-fit $\Lambda$CDM model (lower panel). The black squares with error bars represent the joint lensing data points, plotted alongside the individual bandpowers from Planck PR4 (light yellow), ACT DR6 (light red), and SPT-3G M2PM (light blue) shown as semi-transparent error boxes. Theoretical curves are displayed using the results from CMB+DESI+DES-Dovekie+KiDS-Legacy for the $\Lambda$CDM (solid gray), $w_0w_a$CDM (dotted pink), $\mu_0\Sigma_0\Lambda$CDM (dashed blue), and $\mu_0\Sigma_0 w_0w_a$ (dash-dotted orange) models.}
\end{figure*}

In this section, we report the constraints derived from the combination of CMB, DESI, DES-Dovekie, and KiDS-Legacy datasets, which are listed in Table~\ref{table1}. The two-dimensional contour plots and whisker plots for the two key MG parameters, $\mu_0$ and $\Sigma_0$, are displayed in Figs.~\ref{fig1} and \ref{fig2}, respectively. Figure~\ref{fig3} illustrates the fit to the CMB lensing data. Finally, the model comparison results are summarized in Table~\ref{table2}.

In Fig.~\ref{fig1}, we present the two-dimensional contour plots of $\mu_0$ and $\Sigma_0$ in the left and right panels for the $\mu_0\Sigma_0\Lambda$CDM and $\mu_0\Sigma_0 w_0w_a$ models, utilizing different combinations of CMB-nl, CMB, DESI, DES-Dovekie, and KiDS-Legacy data. In the $\mu_0\Sigma_0\Lambda$CDM model, using CMB-nl and CMB alone yields relatively weak constraints on both $\mu_0$ and $\Sigma_0$. Nevertheless, the inclusion of CMB lensing data improves the constraint precision on $\Sigma_0$ by approximately 10\%. The addition of the two background expansion measurements, DESI and DES-Dovekie, further slightly enhances the constraint precision on $\Sigma_0$ through parameter degeneracy effects, primarily by improving the measurement precision of the $S_8$ parameter. Regarding the parameter $\mu_0$, which characterizes deviations in matter clustering, its constraints are derived solely from the matter clustering effects reflected in the CMB temperature and polarization power spectra. Consequently, the inclusion of CMB lensing, DESI, and DES-Dovekie data has only a marginal impact on the constraints for this parameter, as shown in the left panel of Fig.~\ref{fig1} and the results in Table~\ref{table1}.

As anticipated, when we further consider the CMB+KiDS-Legacy combination, it is evident that the inclusion of KiDS-Legacy weak lensing data significantly improves the constraint precision for both $\mu_0$ and $\Sigma_0$. Specifically, compared to using CMB data alone, the addition of KiDS-Legacy improves the constraint precision on $\mu_0$ by approximately 60\%, and on $\Sigma_0$ by approximately 43\%, as shown in the left panel of Fig.~\ref{fig1}. Meanwhile, the constraints on $\mu_0$ and $\Sigma_0$ from CMB+KiDS-Legacy are also significantly stronger than those from the combination of Planck CMB temperature and polarization and DESY3 data in Ref.~\cite{Ishak:2024jhs}. This further demonstrates that the complementarity between new-generation weak lensing data and CMB data across high and low redshifts effectively breaks parameter degeneracies, highlighting their significant advantage in testing GR. It is worth noting here that due to the discrepancy in $S_8$ measurements between CMB lensing and DESY3, the analysis in Ref.~\cite{Ishak:2024jhs} did not include CMB lensing data when considering DESY3. In this work, however, as mentioned previously, the $S_8$ measurements from CMB and KiDS-Legacy are nearly consistent, allowing for their safe combination. Finally, by combining all datasets (CMB+DESI+DES-Dovekie+KiDS-Legacy), we obtain the most precise constraints in this work: $\mu_0 = 0.21 \pm 0.21$ and $\Sigma_0 = 0.149 \pm 0.051$. It is observed that even without using the FS galaxy clustering data from DESI, we achieve results with even higher precision for $\mu_0$ compared to the DESI(FS+BAO)+CMB-nl+DESY5+DESY3 combination in Ref.~\cite{Ishak:2024jhs}, while the constraint precision for $\Sigma_0$ is comparable. This demonstrates the significant capability of the latest high-precision CMB and KiDS-Legacy data in testing GR.

When considering the background evolution as the $w_0w_a\mathrm{CDM}$ model, unlike in the $\Lambda$CDM case, the inclusion of the two background datasets, DESI and DES-Dovekie, has a more significant impact. This is because they provide more precise constraints on the DE EoS, which in turn imposes stricter constraints on the MG parameters. Specifically, as shown in Table~\ref{table1}, although using only CMB+KiDS-Legacy can constrain $\mu_0$ and $\Sigma_0$ relatively precisely, the insensitivity to background evolution leads to significantly weakened constraints on the DE EoS parameters, $H_0$, and $S_8$; in particular, it fails to provide an effective central value for $S_8$. In contrast, using only CMB+DESI+DES-Dovekie data measures the background evolution relatively accurately but provides weaker constraints on $\mu_0$ and $\Sigma_0$, as shown in the right panel of Fig.~\ref{fig1}. Therefore, combining all data (CMB+DESI+DES-Dovekie+KiDS-Legacy), we simultaneously obtain precise constraints on both the background evolution and the MG parameters. For the DE EoS parameters, we obtain $w_0 = -0.827 \pm 0.055$ and $w_a = -0.59 \pm 0.21$. Simultaneously, for the MG parameters, we obtain $\mu_0 = 0.08^{+0.29}_{-0.24}$ and $\Sigma_0 = 0.115 \pm 0.053$. It is observed that although the introduction of MG parameters at the perturbation level acts to weaken the evidence for dynamical DE, the significance of such evidence persists.

To clearly illustrate the deviations of the MG parameters, we present whisker plots of $\mu_0$ and $\Sigma_0$ utilizing different data combinations and models in Fig.~\ref{fig2}. For the parameter $\mu_0$, the results are generally consistent with the predictions of GR across all datasets and background evolutions. Specifically, due to larger uncertainties in the CMB-nl, CMB, and CMB+DESI+DES-Dovekie combinations, the deviation of $\mu_0$ from zero is only around $0.2\sigma$ level. Although the inclusion of KiDS-Legacy significantly improves the constraints, the central value remains unshifted. It is worth noting that under the $\Lambda$CDM background evolution, the combination of all datasets yields a $1.0\sigma$ deviation for $\mu_0$, whereas this deviation decreases to $0.3\sigma$ in the $w_0w_a\mathrm{CDM}$ background. This suggests that the $1.0\sigma$ deviation may stem from assuming a $\Lambda$CDM background expansion history that is inconsistent with current data. In summary, there is no evidence of deviation from GR in matter clustering at cosmological scales.

Interestingly, for the parameter $\Sigma_0$, deviations from GR exceeding the $1\sigma$ level appear across different data combinations and background evolutions, reaching up to the $3.0\sigma$ level. Specifically, even with relatively weak constraints, the CMB-nl, CMB, and CMB+DESI+DES-Dovekie combinations yield deviations ranging from $1.2\sigma$ to $1.6\sigma$ level. The CMB+KiDS-Legacy combination shows a $2.5\sigma$ deviation in $\mu_0\Sigma_0\Lambda$CDM, which decreases to $1.5\sigma$ in $\mu_0\Sigma_0w_0w_a$. Finally, the most precise data combination yields a deviation as high as $3.0\sigma$ level\footnote{The reconstruction presented in Ref.~\cite{Stolzner:2025vet} within Horndeski gravity model utilizing a combination of KiDS-Legacy data, eBOSS redshift space distortion (RSD), and Planck CMB anisotropy data revealed that $\mu_0$ deviates from GR at approximately $2.9\sigma$ level while no evidence of deviation was found for $\Sigma_0$. We attribute this discrepancy to the differences in the adopted datasets. The variance in the $\Sigma_0$ constraint likely originates from our utilization of CMB anisotropy and lensing data from Planck, ACT, and SPT rather than relying solely on Planck data. The difference in the $\mu_0$ constraint may stem from the absence of galaxy clustering measurements such as RSD data in our analysis.}, and even in $\mu_0\Sigma_0w_0w_a$, it remains at $2.2\sigma$ level. A detailed discussion on how the dynamical dark energy background alleviates this deviation via parameter degeneracies is provided in Appendix~\ref{appendixB}. Previous studies suggested that the deviation of $\Sigma_0$ from the GR value primarily originated from CMB lensing anomalies in Planck PR3 data rather than new physics mechanisms~\cite{Ishak:2024jhs,Renzi:2017cbg,Mokeddem:2022bxa}. This deviation consistently disappeared when using Planck PR4 lensing, weak lensing data (DESY3), and considering various parameterizations~\cite{Ishak:2024jhs}. However, our study reveals re-emerging possible signs of $\Sigma_0$ deviation from the GR value, using the currently most precise and consistent CMB anisotropy, CMB lensing, and weak lensing data.

To intuitively investigate how this lensing deviation is driven by the data, in Fig.~\ref{fig3} we plot the theoretical curves of the CMB lensing power spectrum for the four models, the CMB lensing data, and the residuals relative to $\Lambda$CDM, using the fit results from CMB+DESI+DES-Dovekie+KiDS-Legacy as a baseline. The lensing bandpower data from Planck PR4, ACT DR6, and SPT-3G M2PM are represented by light yellow, light pink, and light blue shaded regions, respectively, while the joint reconstruction lensing data points based on these three are shown as open black error bars. Benefiting from space-based observations free from atmospheric interference, Planck PR4 measurement bandpowers are concentrated at $\ell<400$, dominating the precision at low $\ell$ and providing baseline large-scale information. The SPT-3G M2PM bandpowers cover the widest range of multipoles $\ell$; although measurements are less precise at low $\ell$, it possesses extremely high angular resolution and low noise, thus providing data with very high signal-to-noise ratios at small scales (high $\ell$). The ACT DR6 bandpowers cover relatively intermediate scales and achieve relatively precise measurements across all scales. The complementarity of these three datasets forms the most precise current CMB lensing measurement covering full scales.

Next, we evaluate the performance of different models in fitting the CMB lensing data. The CMB lensing power spectra based on $\Lambda$CDM and $w_0w_a\mathrm{CDM}$ models nearly overlap; considering dynamical DE only slightly raises the amplitude of the power spectrum by $\sim$2\% across full scales, indicating that solely modifying the background cannot significantly alter the lensing power spectrum. Notably, the $\Lambda$CDM and $w_0w_a\mathrm{CDM}$ models provide good fits for the lensing data points at the largest scale ($\ell=27$) and at $\ell>100$, but show large deviations for the two points at $\ell=53$ and $\ell=83.5$, struggling to fit the large-scale lensing data. In contrast, in the $\mu_0\Sigma_0\Lambda$CDM and $\mu_0\Sigma_0 w_0w_a$ models, considering MG effects that deviate from GR significantly raises the amplitude of lensing power spectrum by 5--15\% at larger scales ($\ell<100$) compared to $\Lambda$CDM, especially at lower $\ell$. Consequently, although it deviates from the $\ell=27$ data point, considering MG effects provides a better fit for the points at $\ell=53$ and $\ell=83.5$. This implies that the significant deviation of $\Sigma_0$ from GR may be partially driven by these two CMB lensing points. 

To investigate whether this deviation is solely induced by these two specific data points and to rule out the presence of highly localized systematic errors, we perform a data ablation test. After excluding the bandpowers within the multipole range $\ell \in [40, 100]$, we obtain $\mu_0 = 0.21\pm 0.22$ and $\Sigma_0 = 0.133 \pm 0.056$. This result indicates that $\mu_0$ remains largely consistent with our baseline results, while $\Sigma_0$ still exhibits a deviation of approximately $2.4\sigma$. This demonstrates that while these two data points contribute to elevating $\Sigma_0$ by increasing the statistical significance from $2.4\sigma$ to $3.0\sigma$, the physical preference for $\Sigma_0 > 0$ is fundamentally driven by a broader large-scale power excess. The large-scale measurements from ACT and SPT may require a higher lensing amplitude at relatively large scales, leading to the deviation of the $\Sigma_0$ parameter from GR. Coupled with the stringent constraining power of the KiDS-Legacy data, this underlying excess ultimately establishes the robustness of the observed deviation.

\begin{table}[!htb]
\renewcommand\arraystretch{1.5}
\centering
\caption{The values of $\Delta\chi^2$ and $\Delta\mathrm{DIC}$ for the $w_0w_a\mathrm{CDM}$, $\mu_0\Sigma_0\Lambda$CDM, and $\mu_0\Sigma_0 w_0w_a$ models relative to $\Lambda$CDM. Negative values of $\Delta\chi^2$ and $\Delta\mathrm{DIC}$ indicate a better fit compared to $\Lambda$CDM.}
\label{table2}
\resizebox{0.5\textwidth}{!}{
\setlength{\tabcolsep}{5pt}
\begin{tabular}{lcc}
\hline\hline
Model/Data & $\Delta \chi^2$ & $\Delta$DIC \\
\hline

\multicolumn{3}{l}{$\boldsymbol{w_0w_a\mathrm{CDM}}$} \\
CMB+DESI+DES-Dovekie             & $-14.24$ & $-12.22$ \\
CMB+DESI+DES-Dovekie+KiDS-Legacy & $-13.96$ & $-12.64$ \\
\hline

\multicolumn{3}{l}{$\boldsymbol{\mu_0\Sigma_0\Lambda\mathrm{CDM}}$} \\
CMB+DESI+DES-Dovekie             & $-11.47$ & $-4.55$ \\
CMB+DESI+DES-Dovekie+KiDS-Legacy & $-11.28$ & $-6.16$ \\
\hline

\multicolumn{3}{l}{$\boldsymbol{\mu_0\Sigma_0w_0w_a}$} \\
CMB+DESI+DES-Dovekie             & $-19.59$ & $-10.35$ \\
CMB+DESI+DES-Dovekie+KiDS-Legacy & $-18.10$ & $-14.36$ \\
\hline
\end{tabular}
}
\end{table}

Finally, to quantitatively compare the goodness of fit among different models, we calculate the $\Delta\chi^2$ values relative to the $\Lambda$CDM model and employ the Deviance Information Criterion (DIC) for supplementary model assessment, as presented in Table~\ref{table2}. We find that when simultaneously accounting for deviations at both the background and perturbation levels (i.e., dynamical DE and MG), the $\mu_0\Sigma_0 w_0w_a$ model exhibits the most significant improvement in fit compared to $\Lambda$CDM. This is particularly notable when using the CMB+DESI+DES-Dovekie+KiDS-Legacy dataset, which yields $\Delta\chi^2 = -18.10$ and $\Delta\mathrm{DIC} = -14.36$. On one hand, from the background perspective, this further emphasizes that the recent preference for dynamical DE indicated by DESI persists even within a MG context; this is likely driven primarily by the DESI BAO and DES-Dovekie SN measurements of the background expansion history. On the other hand, at the perturbation level, current data also support indications of deviations from GR, likely stemming mainly from high-precision lensing data provided by CMB lensing and KiDS-Legacy.

In summary, in addition to deviations from the $\Lambda$CDM model in background evolution, current data also indicate signs of deviation in gravitational light deflection from GR at the perturbation level. Whether this phenomenon arises from new physics beyond GR or stems from systematic errors within the data remains to be clarified in future research.

\section{Conclusion}\label{sec4}

In this work, utilizing the latest CMB anisotropy and lensing data, together with the most precise weak gravitational lensing data from KiDS-Legacy, and combining background expansion data from DESI DR2 BAO and DES-Dovekie SN, we have tested deviations from GR at cosmological scales. We considered both $\Lambda$CDM and $w_0w_a\mathrm{CDM}$ background evolutions and adopted a specific parameterization for the time evolution of MG. We report the test results and provide a detailed analysis of the causes of the observed deviations, emphasizing the synergistic complementarity between CMB data and KiDS-Legacy in testing GR at cosmological scales.

We find that the inclusion of KiDS-Legacy significantly improves the measurement precision of MG parameters by approximately 60\% for $\mu_0$ and 43\% for $\Sigma_0$, indicating that the combination of high- and low-redshift data offers effective complementarity and significantly breaks parameter degeneracies. In the $\mu_0\Sigma_0\Lambda$CDM model, using the full dataset combination (CMB+DESI+DES-Dovekie+KiDS-Legacy), we obtain $\mu_0 = 0.21\pm 0.21$ and $\Sigma_0 = 0.149\pm 0.051$. Considering the $\mu_0\Sigma_0 w_0w_a$ model, we can simultaneously impose precise constraints on the DE EoS and MG parameters using the full dataset, yielding $\mu_0 = 0.08^{+0.29}_{-0.24}$, $\Sigma_0 = 0.115\pm 0.053$, $w_0 = -0.827\pm 0.055$, and $w_a = -0.59\pm 0.21$. Crucially, we find that the parameter $\mu_0$ shows no deviation from the GR value across all data combinations and background evolutions, indicating that the clustering of non-relativistic matter is consistent with GR predictions given the current data. However, the parameter $\Sigma_0$ consistently exhibits a deviation from GR, reaching up to the $3.0\sigma$ level. Our analysis suggests that this deviation is primarily driven by a broad-band power excess in the large-scale CMB lensing measurements from ACT and SPT. While the specific data points at $\ell=53$ and $\ell=83.5$ exhibit the most prominent localized excess and enhance the statistical significance, data ablation tests confirm that the preference for $\Sigma_0 > 0$ persists at a $2.4\sigma$ level even when these specific bins are completely removed. This highlights that the deviation is not an artifact of localized systematics, but a robust feature of the combined datasets.

In summary, our study underscores the pivotal role of the synergy between high-precision CMB lensing and weak gravitational lensing data in testing GR at cosmological scales. We reveal that at cosmological scales, the gravitational deflection of light ($\Sigma_0$) appears to exhibit a stronger amplitude than predicted by GR, whereas the clustering of non-relativistic matter ($\mu_0$) remains consistent with GR predictions. In the future, with the release of full-shape power spectrum data from the complete 5-year DESI observations, as well as more precise observational data from Euclid~\cite{Euclid:2024yrr} and Large Synoptic Survey Telescope~\cite{LSST:2008ijt}, we will be able to test GR with higher precision and further determine whether this deviation arises from new physics beyond GR or from systematic errors within the data itself.

\begin{table*}[htbp]
\renewcommand\arraystretch{1.5}
\centering
\caption{Cosmological parameter constraints for the $\mu_0 \Sigma_0 \Lambda$CDM, $\mu_0 \Sigma_0 \Lambda$CDM+$A_\mathrm{lens}$, and $\mu_0 \Sigma_0 \Lambda$CDM+$A_\mathrm{lens}$+$\sum m_\nu$ models, utilizing CMB+DESI+DES-Dovekie+KiDS-Legacy data. Here, $H_0$ is in units of $\rm km\,s^{-1}\,Mpc^{-1}$.}
\label{table3}
\resizebox{\textwidth}{!}{
\setlength{\tabcolsep}{5pt}
\begin{tabular}{lcccccc}
\hline\hline
Data / Model & $H_0$ & $S_8$ & $\mu_0$ & $\Sigma_0$ & $A_\mathrm{lens}$ &  $\sum m_\nu$ [eV]\\
\hline
\multicolumn{7}{l}{\textbf{CMB+DESI+DES-Dovekie+KiDS-Legacy}} \\
$\mu_0\Sigma_0\Lambda\mathrm{CDM}$ & $68.32\pm 0.27$ & $0.826\pm 0.019$ & $0.21\pm 0.21$ & $0.149\pm 0.051$ & $\text{---}$ & $\text{---}$ \\
$\mu_0\Sigma_0\Lambda\mathrm{CDM}+A_\mathrm{lens}$ & $68.32\pm 0.26$ & $0.825\pm 0.020$ & $0.20\pm 0.21$ & $0.120\pm 0.056$ & $1.034\pm 0.029$ & $\text{---}$ \\
$\mu_0\Sigma_0\Lambda\mathrm{CDM}+A_\mathrm{lens}+\sum m_\nu$ & $68.34\pm 0.30$ & $0.824\pm 0.020$ & $0.15\pm 0.23$ & $0.107^{+0.054}_{-0.063}$ & $1.035\pm 0.030$ & $<0.098\,(2\sigma)$ \\
\hline
\end{tabular}
}
\end{table*}

\section*{Acknowledgments}
This work was supported by the National Natural Science Foundation of China (Grants Nos. 12533001, 12473001, and 12575049), the National SKA Program of China (Grants Nos. 2022SKA0110200 and 2022SKA0110203), the China Manned Space Program (Grant No. CMS-CSST-2025-A02), and the National 111 Project (Grant No. B16009).

\section*{Conflict of Interest}
The authors declare that they have no conflict of interest.

\appendix
\section{Impact of CMB lensing amplitude and neutrino mass}\label{appendixA}

\begin{figure}[!htbp]
\includegraphics[scale=0.5]{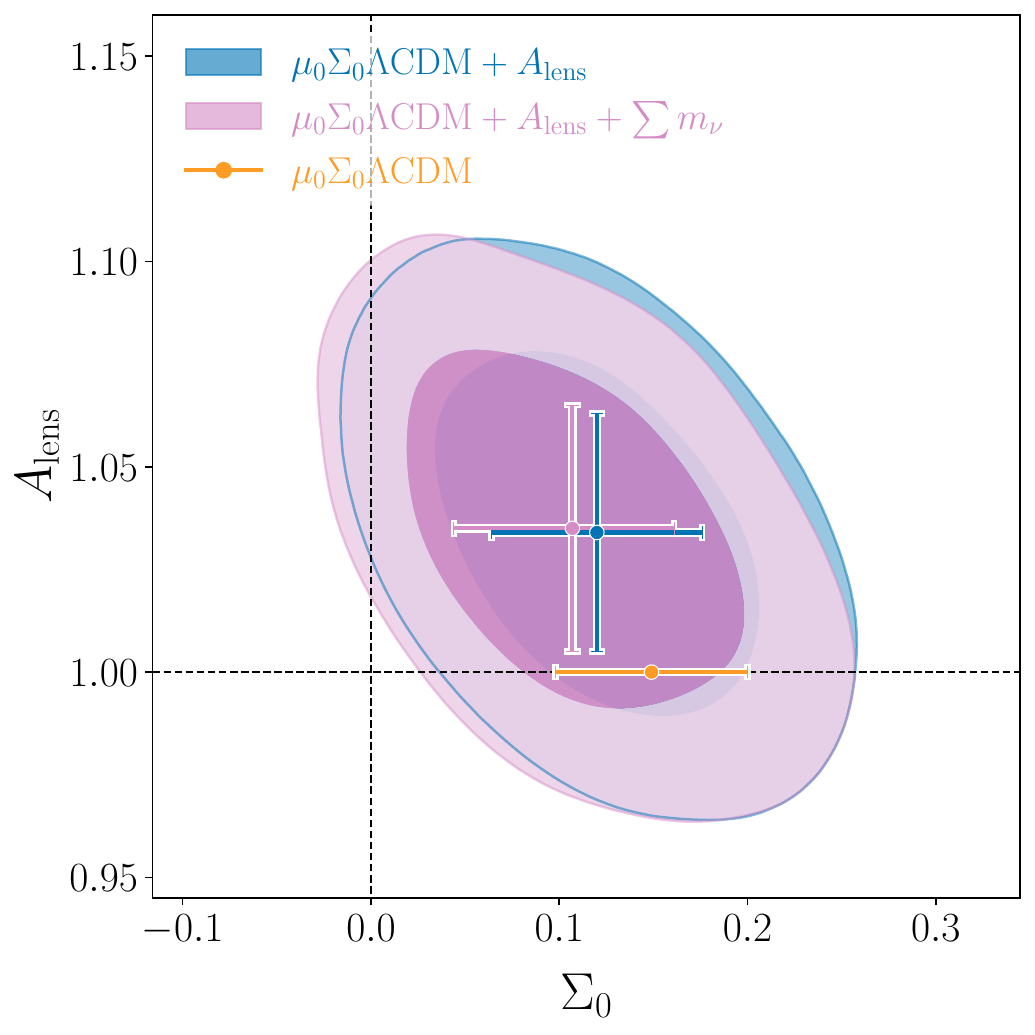}
\centering
\caption{\label{fig4} The $1\sigma$ and $2\sigma$ credible-interval contours for the $A_\mathrm{lens}$ and $\Sigma_0$ parameters, utilizing the CMB+DESI+DES-Dovekie+KiDS-Legacy data.}
\end{figure}

In this appendix, we investigate the influence of the phenomenological CMB lensing amplitude parameter $A_\mathrm{lens}$ and the total neutrino mass $\sum m_\nu$ on the constraints of MG parameters. We extend the baseline $\mu_0\Sigma_0\Lambda$CDM model by treating $A_\mathrm{lens}$ as well as the combination of $A_\mathrm{lens}$ and $\sum m_\nu$ as free parameters. Table~\ref{table3} summarizes the corresponding parameter constraints derived from the joint CMB+DESI+DES-Dovekie+KiDS-Legacy dataset. When the lensing amplitude is allowed to vary freely, we obtain $A_\mathrm{lens} = 1.034 \pm 0.029$, which indicates that the data slightly favor an enhanced lensing effect compared to the standard prediction. As illustrated in Fig.~\ref{fig4}, a negative degeneracy exists between $A_\mathrm{lens}$ and $\Sigma_0$. Consequently, the introduction of the $A_\mathrm{lens}$ parameter absorbs a significant proportion of the large-scale power excess in the lensing data and the gravitational light deflection parameter decreases to $0.120 \pm 0.056$. The statistical significance of the deviation of $\Sigma_0$ from GR accordingly diminishes from approximately $3.0\sigma$ to $2.1\sigma$.

Furthermore, we explore the scenario where the total neutrino mass is simultaneously varied. In the $\mu_0\Sigma_0\Lambda\mathrm{CDM}+A_\mathrm{lens}+\sum m_\nu$ model, we obtain a stringent upper limit of $\sum m_\nu < 0.098$ eV at the $2\sigma$ confidence level. Meanwhile, the gravitational light deflection parameter is measured as $\Sigma_0 = 0.107^{+0.054}_{-0.063}$ and this further reduces the evidence for deviation from GR to the $1.7\sigma$ level. These results demonstrate that the deviation of $\Sigma_0$ reported in our primary analysis is partially degenerate with the phenomenological lensing amplitude and the total neutrino mass.

\section{Impact of a dynamical DE background}\label{appendixB}

\begin{figure*}[!htbp]
\includegraphics[scale=0.5]{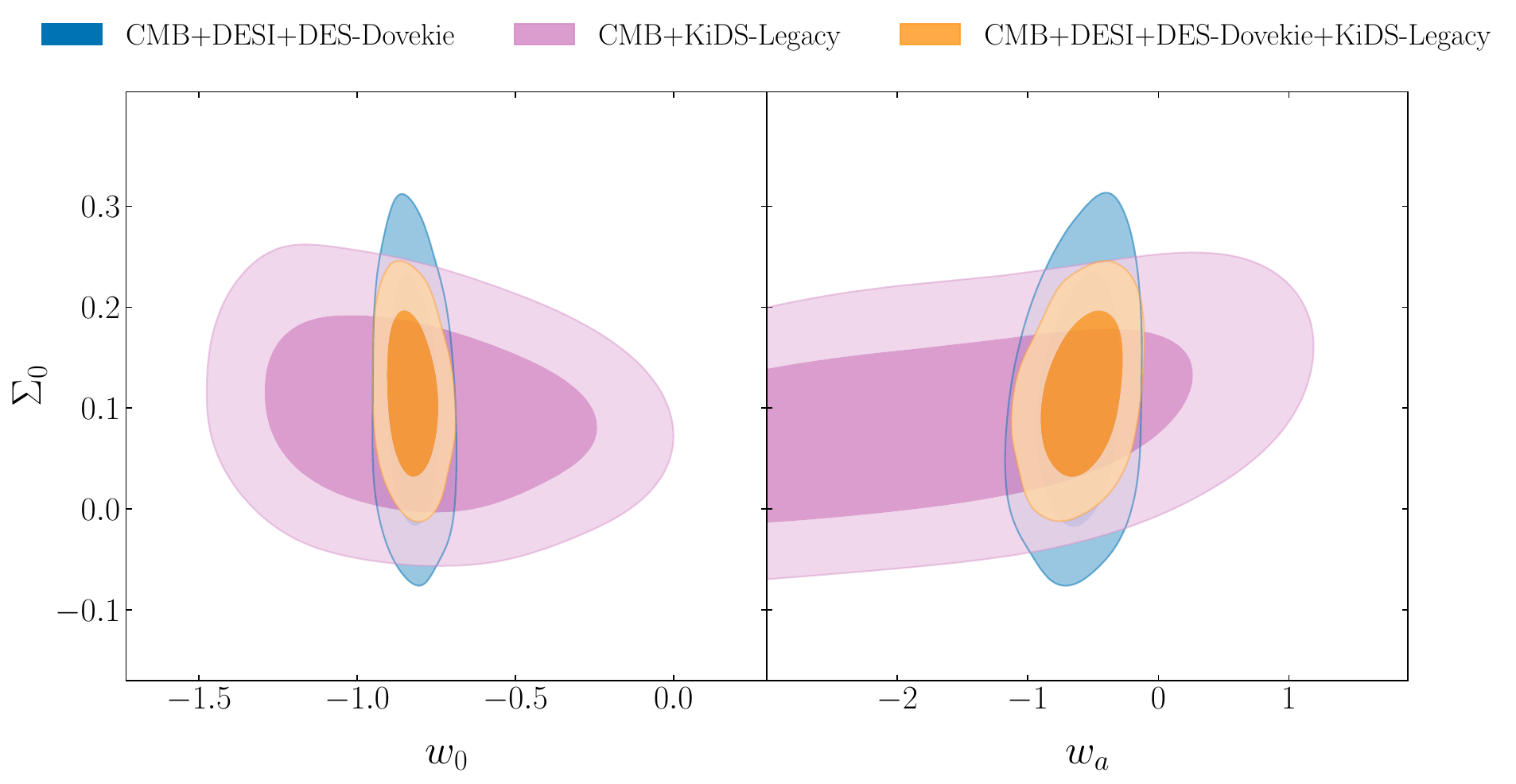}
\centering
\caption{\label{fig5} The $1\sigma$ and $2\sigma$ credible-interval contours for $\Sigma_0$ with $w_0$ and $w_a$ using the CMB+DESI+DES-Dovekie+KiDS-Legacy data.}
\end{figure*}

In this appendix, we further elucidate the influence of dynamical DE on the deviation of MG parameters. Figure~\ref{fig5} illustrates the $1\sigma$ and $2\sigma$ credible-interval contours of $\Sigma_0$ with $w_0$ and $w_a$. The results demonstrate a mild negative degeneracy between $\Sigma_0$ and $w_0$ alongside a mild positive degeneracy between $\Sigma_0$ and $w_a$. The current joint observational dataset exhibits a preference for a dynamical DE background characterized by $w_0 > -1$ and $w_a < 0$. Given the aforementioned degeneracy directions, this specific background evolution naturally drives the $\Sigma_0$ parameter towards lower values. Consequently, the introduction of a dynamical DE background absorbs a fraction of the anomalous lensing signal. These findings explain through the perspective of parameter degeneracy why the statistical significance of the deviation of $\Sigma_0$ from GR decreases from $3.0\sigma$ in the $\mu_0\Sigma_0\Lambda$CDM model to $2.2\sigma$ in the $\mu_0\Sigma_0w_0w_a$ model.

\bibliography{main}

\end{document}